\documentclass[11pt]{article}

\usepackage[a4paper,margin=1in]{geometry}
\usepackage[T1]{fontenc}
\usepackage[utf8]{inputenc}
\usepackage{lmodern}
\usepackage{microtype}
\usepackage{amsmath}
\usepackage{amssymb}
\usepackage{bm}
\usepackage{booktabs}
\usepackage{graphicx}
\usepackage{subcaption}
\usepackage{xcolor}
\usepackage[numbers,sort&compress]{natbib}
\usepackage[colorlinks=true,linkcolor=blue!50!black,citecolor=blue!50!black,urlcolor=blue!50!black]{hyperref}
\usepackage[nameinlink,noabbrev]{cleveref}

\graphicspath{{figures/}}

\newcommand{\dd}{\mathrm{d}}
\newcommand{\GeV}{\mathrm{GeV}}
\newcommand{\TeV}{\mathrm{TeV}}
\newcommand{\pT}{p_{\mathrm{T}}}

\newcommand{\vegas}{\textsc{Vegas}}
\newsavebox{\crosssectionsteinplotbox}

\title{Event Generation with Parallel Langevin Sampling and Learned Stein Diagnostics}
\author{
  \parbox{0.96\textwidth}{\centering
    Rob Verheyen\footnote{\href{mailto:robverheyen@gmail.com}{robverheyen@gmail.com}}\\
    {\small\textit{Department of Physics and Astronomy, University College London, London, WC1E 6BT, UK}}
  }
}
\date{}

\begin{document}

\maketitle

\begin{abstract}
Efficient event generation is a major computational challenge for precision collider phenomenology, especially for high-multiplicity final states where matrix-element evaluations are expensive and rejection-sampling efficiencies are low.
We study an alternative approach based on many parallel underdamped Langevin chains, retaining one terminal state from each chain to obtain unweighted events while avoiding within-chain autocorrelation.
A learned Stein discrepancy is used as a convergence diagnostic, providing a data-driven estimate of the relaxation time.
We apply the method to tree-level $u\bar u\to Z+n g$ event generation and find that relaxation requires only a modest number of exact-target Langevin steps, with mild growth over the multiplicities studied.
Finally, we show that simple neural-network surrogate initialization can substantially reduce the required number of exact matrix-element and gradient evaluations.
\end{abstract}

\section{Introduction}
The upcoming high-luminosity upgrade to the Large Hadron Collider (LHC) \cite{ZurbanoFernandez:2020cco} and its potential successors \cite{FCC:2018evy,FCC:2018vvp} are slated to achieve unprecedented experimental precision.
The theory side must match this precision, not only in systematic control but also in statistical accuracy. 
With current technical implementations, the latter, in the form of the required number of Monte Carlo (MC) event samples, is expected to become the dominant computational bottleneck~\cite{Buckley:2011ms,Narain:2022qud,HEPSoftwareFoundation:2017ggl,HSFPhysicsEventGeneratorWG:2020gxw,EuropeanStrategyGroup:2020pow}.

The most expensive component of the MC event generator chain~\cite{vanBeekveld:2026uxl} is the sampling of the hard proton-proton collision differential cross section
\begin{equation} \label{cross-section}
  \dd \sigma_{p p \rightarrow X_n} = \frac{1}{\mathcal{F}} \sum_{i,j} f_i(x_1) \, f_j(x_2) \, |M_{ij \rightarrow X_n}(\Phi_n)|^2 \, \dd x_1 \, \dd x_2 \, \dd\Phi_n,
\end{equation}
where the sum runs over the constituents of the proton (quarks, antiquarks and gluons), $f$ are their corresponding parton density functions, $|M_{ij\rightarrow X_n}|^2$ is the squared matrix element, and $\mathcal{F}$ is the incoming partonic flux factor.
The integration measure is given by the two partonic momentum fractions $\dd x_{1,2} \in [0,1]$ and the $n$-body phase-space measure $\dd\Phi_n$, which integrates over the final-state momenta while ensuring momentum conservation.
The total sampling dimensionality of this object is $3n-2$. 

For large multiplicities $n$ and for high theoretical precision, the evaluation of the matrix element can be prohibitively expensive.
The method used to sample Eq.~\eqref{cross-section} must therefore be as efficient as possible in terms of evaluations, even if this comes at increased algorithmic cost. 
Current methods are still largely based on the \vegas{} algorithm~\cite{Lepage:1977sw,Lepage:2020tgj} with many physics-inspired improvements~\cite{Kleiss:1994qy,Ohl:1998jn,Maltoni:2002qb,Krauss:2001iv,Papadopoulos:2000tt,Kilian:2007gr,Jadach:2002kn,Gleisberg:2008fv}.
These methods all rely on an adaptive approximation of an unnormalized target density $\widetilde{p}(x)$ (the differential cross section) by some density $q(x)$ that is easily sampled. 
One can then sample from $q(x)$ and assign weights $w(x) = \widetilde{p}(x)/q(x)$. 
To obtain unweighted samples, rejection sampling must be performed with acceptance probability 
\begin{equation} \label{unweight-efficiency}
  p_{\texttt{acc}}(x) = \frac{w(x)}{\max_x w(x)}.
\end{equation}
The unweighting efficiency $\epsilon = \mathbb{E}_{x \sim q} \left[p_{\texttt{acc}}(x)\right]$ then determines the required number of matrix element evaluations per generated sample.
Eq.~\ref{unweight-efficiency} is very sensitive to the value of $\max_x w(x)$, which may originate from tiny regions of phase space.
Thus, to improve efficiency, one may make use of a reduced maximum and leave some undesirable weighted events~\cite{Sherpa:2024mfk}. 
Nevertheless, for increasing multiplicities, the unweighting efficiency can still degrade by several orders of magnitude~\cite{Hoche:2019flt}.

Recent work has used modern machine-learning (ML) architectures such as normalizing flows in place of \vegas{} to improve efficiencies~\cite{Klimek:2018mza,Bothmann:2020ywa,Gao:2020zvv,Stienen:2020gns,Heimel:2022wyj,Verheyen:2022tov,Heimel:2023ngj,Heimel:2024wph}.
These approaches are however still ultimately based on rejection sampling and thus inherit its disadvantages. 
In this work, we propose a different approach based on parallelized Markov chain Monte Carlo (MCMC) supplemented by an ML-based convergence diagnostic. 
MCMC methods have previously been explored for hard-process phase-space sampling and event generation~\cite{Kharraziha:1999iw,Weinzierl:2001ny,Kroeninger:2014bwa,LaCagnina:2024wcc}.
Our method places particular emphasis on avoiding within-chain autocorrelations and on reaching MCMC stationarity, i.e. samples being representative of $\widetilde{p}(x)$. 
Furthermore, we investigate how neural network surrogates of the target distribution can further reduce the required number of matrix-element evaluations~\cite{Danziger:2021eeg,Janssen:2023ahv,Herrmann:2025nnz}.\footnote{The code used in this study is available at \url{https://github.com/rbvh/ula_stein_evt_gen}.} 

\section{Method}

We now describe our sampling construction, targeting an unnormalized density $\widetilde{p}(x)$ on the unit hypercube $x\in[0,1]^d$.
To avoid within-chain autocorrelation, we evolve many Markov chains in parallel on hardware accelerators and retain only the terminal state from each chain.
The learned Stein discrepancy introduced below is used to determine the relaxation time $t_{\mathrm{rel}}$, i.e. the number of MCMC steps after which the ensemble is representative of the target.

\subsection{Underdamped Langevin Algorithm}
Our MCMC framework is designed for rapid convergence under two practical constraints: the sampler should be straightforward to run over many parallel chains, and it should use expensive target-density and gradient evaluations sparingly.
Sophisticated Hamiltonian Monte Carlo (HMC) methods~\cite{neal2011mcmc} like NUTS~\cite{hoffman2014no} implement adaptive trajectory lengths which leads to complex control flow and thus difficult parallelizability.
Moreover, HMC typically requires several intermediate gradient evaluations before a single Metropolis-Hastings (MH) correction is applied.
We therefore use an Underdamped Langevin Algorithm (ULA)~\cite{roberts1996exponential,cheng2018underdamped}, which retains momentum between steps while applying an MH correction after each proposal.

Each MCMC chain carries a position $x$ and a velocity $v$. 
We assume we have access to the score $s(x) = \nabla \log \widetilde{p}(x)$. 
For stability purposes the score is regularized component-wise to $\widetilde{s}_i(x) = s_i(x) / (1 + \eta_i |s_i(x)|)$, where $\eta_i$ are the MCMC step sizes~\cite{brosse2019tamed}.
At time step $t$, a new proposal $x'$, $v'$ is generated following the leapfrog step 
\begin{align}
  v_{1/2} &= v_t + \frac{1}{2}\Lambda_\eta \, \widetilde{s}(x_t), \nonumber \\
  x' &= x_t + \Lambda_\eta \, v_{1/2}, \nonumber \\
  v' &= v_{1/2} + \frac{1}{2}\Lambda_\eta \, \widetilde{s}(x'),
\end{align}
where $\Lambda_\eta=\operatorname{diag}(\eta_1,\ldots,\eta_d)$.
This step discretizes the Hamiltonian dynamics on the augmented density
\begin{equation}
  \pi(x,v) \propto \widetilde{p}(x)\exp\left(-\frac{1}{2}v^Tv\right).
\end{equation}
In the continuous-time limit, the Hamiltonian flow conserves the Hamiltonian $H(x,v) = -\log \widetilde{p}(x) + v^Tv/2$ and preserves phase-space volume. 
As a result, $\pi(x,v)$ as well as its $x$-marginal, which is proportional to the desired target density $\widetilde{p}(x)$, is invariant.
However, at finite step sizes an integration error is introduced.
By applying a MH acceptance step with probability
\begin{equation}
  a = \min\left\{1, \exp\left[-H(x',v') + H(x_t,v_t)\right]\right\}
\end{equation}
the discretization bias is corrected.
Upon rejection, the position is kept and the incoming momentum is reversed $v_t \rightarrow -v_t$. 
After the MH step, the momentum is refreshed
\begin{equation}
  v_{t+1} = \beta v_{\mathrm{acc}} + \sqrt{1-\beta^2}\,\zeta, \qquad \zeta \sim \mathcal{N}(0,I).
\end{equation}
These momentum updates preserve the Gaussian velocity marginal and inject randomness into the otherwise deterministic Hamiltonian evolution, allowing the chain to move between trajectories and explore the full target distribution.
After equilibration, terminal positions drawn from these chains are distributed according to $\widetilde{p}(x)$.

We write the step sizes as $\log\eta_i=\log\epsilon+\log\rho_i$, with $\sum_i\log\rho_i=0$.
The scalar scale $\epsilon$ is initialized by doubling or halving until the average acceptance probability is close to one half~\cite{hoffman2014no}, while the directions $\rho_i$ are initialized from the diagonal score second moment, motivated by Fisher preconditioning~\cite{girolami2011riemann,Titsias:2023},
\begin{equation}
  F_{ii,t} = \mathbb{E}_{x_t}\left[s_i(x_t)^2\right],
  \qquad
  \log\rho_i = -\frac{1}{2}\log F_{ii,t} + \frac{1}{2d}\sum_{j=1}^d \log F_{jj,t}.
\end{equation}
This fixes $\prod_i\rho_i=1$ and increases step sizes in directions with smaller typical score variation.
During the chain, $\log\epsilon$ is adapted by ascending the expected squared jump distance $\mathrm{ESJD}(\epsilon)= \mathbb{E}_{x_t,v_t,\zeta}\left[a \lVert x' - x_t \rVert^2\right]$ at fixed $\rho$, a standard proxy for MCMC efficiency~\cite{gelman1997weak,pasarica2010adaptively}.
The directions are adapted by updating $\rho_i$ with new estimates of the diagonal score second moment.
Both the ESJD scale update and the score-moment direction update use the same diminishing Robbins--Monro schedule, so that the transition-kernel changes vanish asymptotically, as required for stationarity~\cite{robbins1951stochastic,Roberts:2007adaptive}.
All expectations are evaluated over the parallel chain ensemble.

\subsection{Mirror Langevin Dynamics}
After the application of a mapping, the phase-space coordinates used for the cross-section integrand live on the bounded domain $x\in[0,1]^d$, while ULA is most naturally formulated on $\mathbb{R}^d$. 
We therefore perform the MCMC evolution in mirror space~\cite{Hsieh:2018mld}, in which Langevin dynamics evaluated through a suitable mirror map retain the convergence properties of the corresponding unconstrained algorithm.
We use the element-wise transform $y_i = \Phi^{-1}(x_i) = \sqrt{2}\,\operatorname{erf}^{-1}(2x_i-1)$.
The mirrored target density is the corresponding change of variables
\begin{equation}
  \log \widetilde{p}_{\mathrm{mir}}(y) = \log \widetilde{p}(\Phi(y)) - \frac{1}{2}\lVert y\rVert^2 + \mathrm{const.}
\end{equation}
Note that the constant drops out in the score. 

\subsection{Learned Stein Discrepancy}
To monitor convergence of the ULA chains, we need a distributional discrepancy that can be evaluated using samples from the chain and the unnormalized density of the target.
The Stein discrepancy provides such a diagnostic~\cite{gorham2015measuring,anastasiou2023stein}.
For a vector-valued critic function $f:\mathbb{R}^d\to\mathbb{R}^d$, the Stein discrepancy between a smooth, sampled distribution $p$ and a smooth target distribution $q$ is
\begin{equation} \label{stein_discrepancy}
  S(p,q) = \sup_{f \in \mathcal{F}} \mathbb{E}_{x\sim p} \left[s_q(x)^T f(x) + \nabla \cdot f(x)\right],
\end{equation}
where $\mathcal{F}$ is a norm-bounded class of differentiable functions for which the expectation over $p$ is finite and $q(x) f(x)$ has vanishing tails.
For sufficiently rich critic classes, this quantity is non-negative and satisfies $S(p,q)=0$ if and only if $p=q$~\cite{stein2004use}.

Kernel Stein discrepancies are a popular choice for $\mathcal{F}$~\cite{liu2016kernelized,chwialkowski2016kernel,gorham2017measuring}, but their standard estimators involve pairwise sums that scale as $n^2$ in the number of samples $n$. 
Since we typically work in a regime with large numbers of samples, $n^2$ evaluation is infeasible, and a more natural choice would be one that exploits this large data size. 
We thus use the learned Stein discrepancy (LSD)~\cite{hu2018stein,grathwohl2020learning}, taking $\mathcal{F}$ to be a class of square-integrable differentiable functions that satisfy the Stein boundary condition, and representing the critic by a neural network $f_\phi$.
In practice, square integrability is enforced by adding a quadratic regularization term to the training objective,
\begin{equation} \label{LSD}
  \mathcal{L}_\lambda(\phi; p,q) = - \mathbb{E}_{x\sim p} \left[s_q(x)^T f_\phi(x) + \nabla \cdot f_\phi(x) - \lambda \lVert f_\phi(x) \rVert^2 \right].
\end{equation}
After training, the learned Stein discrepancy is evaluated with the learned critic on a held-out sample, without the regularization term.

\subsection{Sharp Boundaries}
\label{sharp_boundaries_section}
Physical cross sections typically involve sharp phase-space cuts, which restrict the target density to an allowed region with a hard boundary. 
The derivation of Eq.~\ref{stein_discrepancy} involves an integration by parts over this region, and a boundary contribution will thus remain unless the Stein test function has vanishing flux through the cut surface \cite{liu2022estimating,williams2023approximate}. 
We address this issue by introducing a scalar boundary function $h(x)$ that is positive inside the allowed region and vanishes on the cut boundary.
The Stein operator is then applied to the vector field $h f$ rather than to $f$ itself~\cite{liu2022estimating,xu2022standardisation},
\begin{equation}
  S_h(p,q) = \sup_{f \in \mathcal{F}} \mathbb{E}_{x\sim p} \left[h(x) s_q(x)^T f(x) + h(x) \nabla \cdot f(x) + \nabla h(x)^T f(x)\right].
\end{equation}

\subsection{Surrogates}
The sampling strategy described above relies on gradients of the target density.
Differentiable matrix element implementations have recently become available~\cite{Heinrich:2022xfa,Carrazza:2021zug,Carrazza:2021gpx,Heimel:2024wph,Heimel:2026hgp}, enabling the sampling strategy set out above, but at higher computational expense. 
This motivates us to consider the use of learned surrogates to reduce the required number of evaluations.

Neural surrogates for scattering amplitudes and matrix elements have been studied extensively~\cite{Bishara:2019iwh,Badger:2020uow,Aylett-Bullock:2021hmo,Maitre:2021uaa,Danziger:2021eeg,Badger:2022hwf,Janssen:2023ahv,Maitre:2023dqz,Bahl:2024gyt,Brehmer:2024yqw,Breso-Pla:2024pda,Herrmann:2025nnz,Bahl:2025xvx,Favaro:2025pgz,Villadamigo:2025our,Beccatini:2025tpk,Bahl:2026jvt,Bahl:2026qaf}.
In the present setting, one can train a surrogate on a relatively limited set of points without requiring gradients.
Then, chains can be evolved cheaply through the early part of the trajectory using the surrogate as target, before switching over to a shorter sequence of ULA steps using the exact target. 
The surrogate is thus only used to accelerate equilibration, while the final evolution remains tied to the physical target density.
We investigate this strategy in Section~\ref{sec:results}.

\section{Validation}
Before applying the method to collider cross sections, we validate the Stein diagnostic on a target for which the MCMC convergence can be computed analytically.
Starting from a fixed point $\mu_0$ away from some target Gaussian mean $\mu_q$ and diagonal target $\Sigma_q$, we sample an exact Ornstein--Uhlenbeck (OU) transition kernel
\begin{equation}
  x_{t+1} = \mu_q + e^{-\eta \Sigma_q^{-1}}(x_t-\mu_q) + \Sigma_q^{1/2} \left(I-e^{-2\eta\Sigma_q^{-1}}\right)^{1/2}\xi, \qquad \xi\sim\mathcal{N}(0,I),
\end{equation}
where $\eta$ is the step size.
After $t$ steps, the marginal distribution is Gaussian with
\begin{align}
  \mu_t &= \mu_q + e^{-t\eta\Sigma_q^{-1}}(\mu_0-\mu_q) \nonumber \\
  \Sigma_t &= \Sigma_q\left(I-e^{-2t\eta\Sigma_q^{-1}}\right).
\end{align}
Thus $p_t$ converges to the target density as $t\to\infty$, yielding a controlled and testable sequence of distributions.
To validate the role of the boundary function, we apply a radial cut that removes $20\%$ of the target probability mass centered around the target mean $\mu_q$.

Under Eq.~\eqref{LSD}, the optimal Stein discrepancy is~\cite{hu2018stein}
\begin{equation}
  S_{\mathrm{opt}}(p_t,q) = \frac{1}{2\lambda} \mathbb{E}_{x\sim p_t} \left[ \left\lVert s_q(x)-s_{p_t}(x) \right\rVert^2 \right].
\end{equation}
We compare this optimal value against the LSD computed on OU samples and on ULA samples. 
The validation results use dimensions $d=4$ and $d=16$, with target parameters
\begin{equation}
  \mu_q = \operatorname{linspace}(-2,2,d),
  \qquad
  \Sigma_q = \operatorname{diag}\!\left(\operatorname{linspace}(0.5,2,d)^2\right).
\end{equation}
The OU process is initialized from the fixed point $\mu_0=\mu_q+3$, while ULA is initialized from a unit Gaussian.
The LSD regularization is $\lambda=0.1$.
The Stein critic is modelled with a five-layer residual GeLU network with hidden width 128.
Optimization is performed with the Adam optimizer~\cite{kingma2014adam} on $n=10^7$ chains, and final LSD values are obtained using 10-fold cross validation.
The boundary function is $h(x) = \max\{1-r_{\mathrm{cut}}^2 / r^2(x),0\}$.

\begin{figure}
\includegraphics[width=\textwidth]{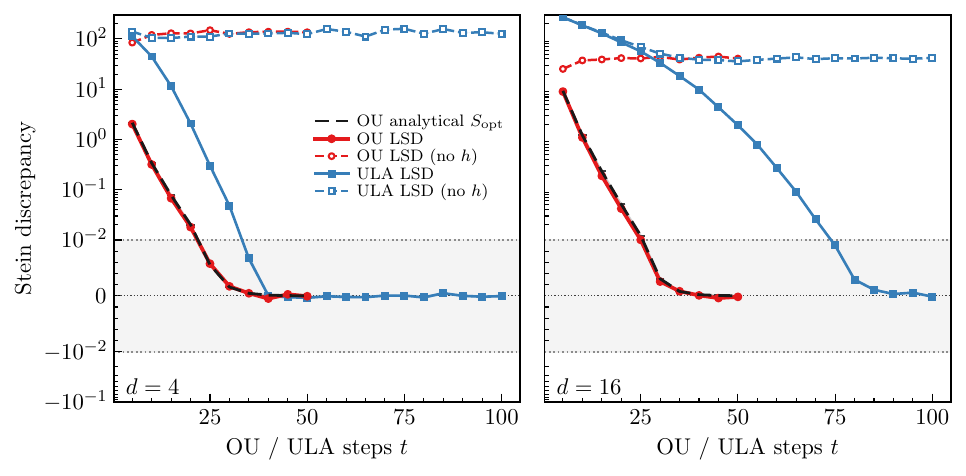}
\caption{The LSD as a function of the number of steps for the exact OU reference process with $\eta=0.5$ (red) and the underdamped Langevin algorithm (ULA, blue). Dashed lines show the boundary-function ablation. The black line shows the analytic optimal Stein discrepancy $S_{\mathrm{opt}}$.}
\label{OU_figure}
\end{figure}

The validation results are shown in Fig.~\ref{OU_figure}.
For the exact OU samples, the learned LSD tracks the analytic value $S_{\mathrm{opt}}$, supporting its use as a quantitative convergence diagnostic. 
The ULA curves show similar monotonic decay towards zero, with the higher-dimensional case converging more slowly, qualitatively consistent with the expected $\sqrt{d}$ scaling of Metropolis-adjusted Langevin methods \cite{chewi2021optimal}. 
In the absence of the boundary function, the remaining finite boundary contribution prevents the Stein discrepancy from approaching zero.

\section{Results} \label{sec:results}

We now apply ULA and the LSD diagnostic to sample the cross section for the process $u \bar{u} \rightarrow Z + n g$.
Matrix elements and their gradients are evaluated using MadJax~\cite{Heinrich:2022xfa} with several modifications for numerical stability. 
We use the NNPDF4.0 LO set~\cite{NNPDF:2021njg}, and implement the PDF and $\alpha_s$ interpolation in JAX~\cite{bradbury2018jax} following the LHAPDF interpolation prescription~\cite{Buckley:2014ana}.
We set $\sqrt{s}=13 \,\TeV$ and apply cuts $\pT>20 \, \GeV$ and $|y|<5$ to all final-state particles and a separation cut $\Delta R>0.4$ to the gluons. 
Phase space is mapped to the $[0,1]^{3n+1}$ hypercube using the Chili~\cite{Bothmann:2023siu} vector boson plus jets map. 
The Stein diagnostic uses the same critic architecture and optimization strategy as in the OU validation, with $\lambda=0.1$.
The boundary function is
\begin{equation}
  h(x) = \left[ \frac{1}{2+\binom{n}{2}} \left( \frac{1}{1-x_a} + \frac{1}{1-x_b} + \sum_{1\leq i<j\leq n} \frac{1}{1-\Delta R_{\mathrm{cut}}^2 / \Delta R_{ij}^2}\right)\right]^{-1},
\end{equation}
for points that pass the cuts, and $0$ for points that do not. 

\begin{figure}
  \centering
  \sbox{\crosssectionsteinplotbox}{\includegraphics[width=0.54\textwidth]{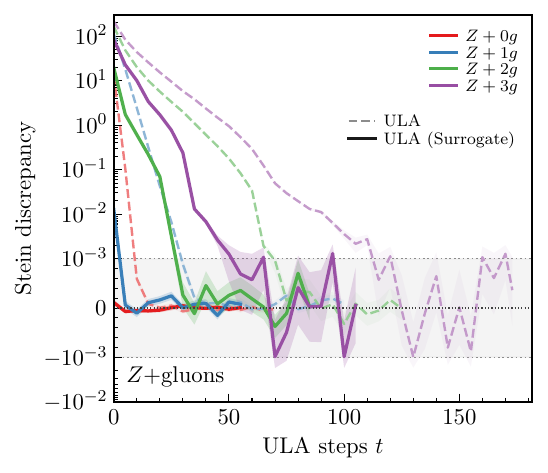}}
  \begin{minipage}[t]{0.54\textwidth}
    \centering
    \vspace{0pt}
    \usebox{\crosssectionsteinplotbox}
    \caption{The LSD as a function of the number of ULA steps for $Z+n g$ cross sections. Colours denote gluon multiplicities, while line styles distinguish ULA with and without surrogate initialization.}
    \label{cross_section_stein_figure}
  \end{minipage}
  \hfill
  \begin{minipage}[t]{0.40\textwidth}
    \centering
    \vspace{0pt}
    \small
    \begin{minipage}[c][\dimexpr\ht\crosssectionsteinplotbox+\dp\crosssectionsteinplotbox\relax][c]{\textwidth}
      \centering
      \begin{tabular}{lcc}
        \toprule
        Process & ULA & ULA (Surrogate) \\
        \midrule
        $Z+0g$ & 20 & 5 \\
        $Z+1g$ & 50 & 5 \\
        $Z+2g$ & 75 & 35 \\
        $Z+3g$ & 115 & 55 \\
        \bottomrule
      \end{tabular}
    \end{minipage}
    \captionof{table}{Relaxation-time checkpoints $t_{\mathrm{rel}}$ for the cross-section runs.}
    \label{tab:cross_section_relaxation_times}
  \end{minipage}
\end{figure}

We use ULA and the LSD diagnostic to generate $10^7$ samples for $n = 0,1,2,3$, both with and without neural surrogates. 
The surrogate is modelled with the same architecture as the Stein critic.
It is optimized in log space with the Adam optimizer and a mean-squared error loss using only $10^6$ samples.
We then run ULA on the surrogate for a large number of steps. 
The terminal states are then used as the initial states for the ULA run, which is otherwise initialized with unit Gaussian samples.
We define the relaxation time $t_{\mathrm{rel}}$ as the first checkpoint for which the learned LSD is no more than one standard error above zero, $S_{\mathrm{LSD}}\leq \Delta S_{\mathrm{LSD}}$.
After the trigger, we evolve the chain for a further $\max(0.5\,t_{\mathrm{rel}},50)$ ULA steps.
The convergence results are shown in Fig.~\ref{cross_section_stein_figure}, and relaxation-time triggers are summarized in Table~\ref{tab:cross_section_relaxation_times}.
The surrogate warm start substantially reduces the number of exact-target ULA steps required for relaxation.

\begin{figure}
\begin{subfigure}{0.49\textwidth}
\includegraphics[width=\textwidth]{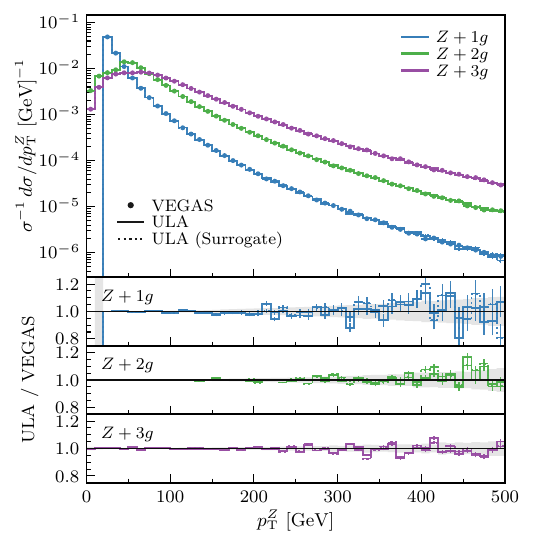}
\end{subfigure}
\hfill
\begin{subfigure}{0.49\textwidth}
\includegraphics[width=\textwidth]{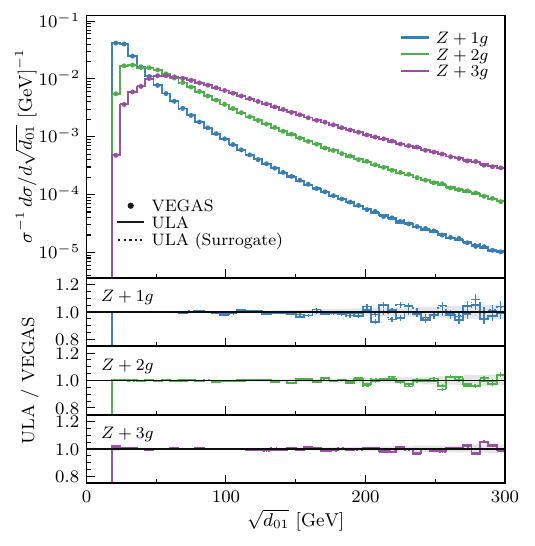}
\end{subfigure}
\caption{Normalized differential distributions for the transverse momentum of the $Z$ boson and the exclusive $k_t$ clustering scale $\sqrt{d_{01}}$. Filled markers show the \vegas{} reference, solid lines show ULA samples, and dotted lines show ULA samples after surrogate initialization. The lower panels show the corresponding ratios to \vegas{} for each multiplicity.}
\label{z_observable_validation_one}
\end{figure}

\begin{figure}
\begin{subfigure}{0.49\textwidth}
\includegraphics[width=\textwidth]{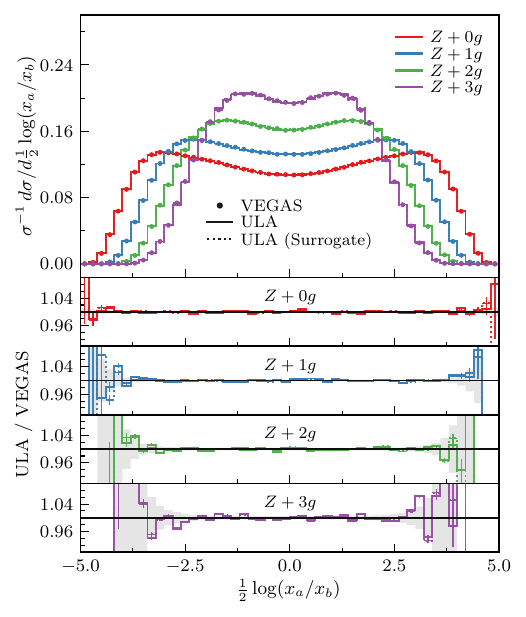}
\end{subfigure}
\hfill
\begin{subfigure}{0.49\textwidth}
\includegraphics[width=\textwidth]{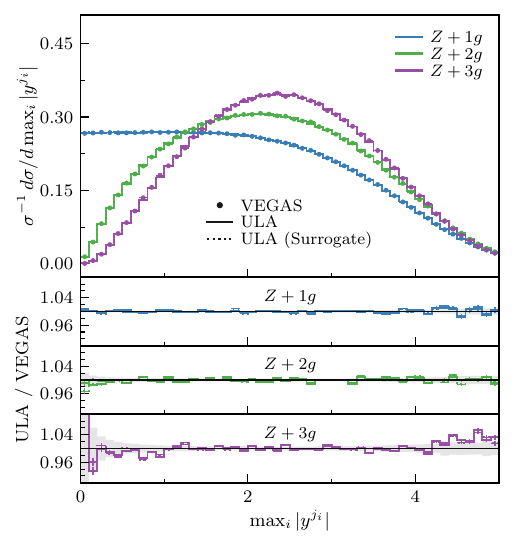}
\end{subfigure}
\caption{Normalized differential distributions for the rapidity of the full final-state system, $\frac{1}{2}\log(x_a/x_b)$, and the largest absolute gluon rapidity, $\max_i |y^{j_i}|$. Filled markers show the \vegas{} reference, solid lines show ULA samples, and dotted lines show ULA samples after surrogate initialization. The lower panels show the corresponding ratios to \vegas{} for each multiplicity.}
\label{z_observable_validation_two}
\end{figure}

To confirm convergence of the ULA chains, we show the normalized distributions of several nontrivial physical observables in Figures~\ref{z_observable_validation_one} and~\ref{z_observable_validation_two}. 
We validate ULA results with and without surrogate initialization against independent \vegas{} estimates. 
Figure~\ref{z_observable_validation_one} shows the vector-boson transverse momentum $p_{\mathrm{T}}^{Z}$ and the first jet-resolution scale $\sqrt{d_{01}}$, obtained by exclusive $k_t$ clustering with radius parameter $R=1$~\cite{Catani:1993hr,Ellis:1993tq}.
Both observables start at $Z+1g$ and their distributions span multiple orders of magnitude. 
Figure~\ref{z_observable_validation_two} shows the longitudinal boost rapidity of the full final-state system, $\frac{1}{2}\log(x_a/x_b)$, and the largest absolute gluon rapidity, $\max_i |y^{j_i}|$.
In all cases, the terminal ULA samples reproduce the \vegas{} distributions within statistical uncertainties. 

\section{Discussion}

The results above suggest that gradient-based parallel MCMC is a practically viable alternative to rejection-based event generation, at least in the regime studied here.
The central practical quantity is the number of exact matrix-element evaluations needed before terminal samples can be used.
For the $Z+n g$ examples, the learned Stein diagnostic triggers after a modest number of exact-target ULA steps, and the increase with multiplicity is mild over the range considered.
This is encouraging because unweighting efficiencies can degrade by orders of magnitude as the final-state multiplicity increases~\cite{Danziger:2021eeg,Janssen:2023ahv}. 
Furthermore, the methods explored here generate strictly unweighted events, while in the usual rejection sampling setup residual weights may remain when using the maximum weight would otherwise reduce the efficiency even further.
The increased algorithmic cost of our method is negligible when compared with the compute required for the matrix-element evaluations.
Like conventional rejection-based generation, our method is also naturally parallel: each chain can be evolved independently, while the adaptation and Stein diagnostic require only ensemble-level reductions.
This makes the method well-matched to GPU/TPU execution.

The required number of evaluations can be further reduced through the use of surrogate initialization.
In our implementation the surrogate is deliberately simple: it uses the same network architecture as the Stein critic and is trained on a comparatively small set of uniformly sampled phase-space points.
Nevertheless, it substantially shortens the exact-target relaxation time.
This should be viewed as a conservative demonstration rather than an optimized surrogate strategy.
More expressive architectures, improved training samples, or active refinement in the regions explored by the chains could plausibly make the warm start closer to the physical target and further reduce the number of exact matrix-element and gradient evaluations.

It is worth noting that the additional computational cost of gradient evaluation is relatively mild.
Reverse-mode automatic differentiation requires only a single forward and backward pass~\cite{baur1983complexity,griewank2012invented}, regardless of the input dimensionality.
At least for tree-level squared matrix elements, which are composed of rational functions of Lorentz-invariant (pseudo)scalars, the computational cost of gradient evaluation should therefore approach only a small constant factor over the value evaluation with sufficiently efficient implementations.

The present study considers a fixed partonic channel, tree-level matrix elements, and a simple phase-space map. 
The observed scaling, and more generally any comparison with current methods, can only be tested in the context of an implementation in full event generator programs.
Still, the combination of parallel chains, an explicit relaxation-time diagnostic, and surrogate initialization provides a concrete route towards event generation in which the expensive exact target is evaluated as few times as possible.

\section*{Acknowledgements}
This work was supported by the European Research Council (ERC) under the European Union’s Horizon 2020 research and innovation programme (grant agreement No. 788223, PanScales).

\bibliographystyle{JHEP}
\bibliography{refs}

@article{Heimel:2022wyj,
  author        = {Heimel, Theo and Winterhalder, Ramon and Butter, Anja and Isaacson, Joshua and Krause, Claudius and Maltoni, Fabio and Mattelaer, Olivier and Plehn, Tilman},
  title         = {{MadNIS - Neural multi-channel importance sampling}},
  eprint        = {2212.06172},
  archivePrefix = {arXiv},
  primaryClass  = {hep-ph},
  doi           = {10.21468/SciPostPhys.15.4.141},
  journal       = {SciPost Phys.},
  volume        = {15},
  number        = {4},
  pages         = {141},
  year          = {2023}
}

@article{Heimel:2023ngj,
  author        = {Heimel, Theo and Huetsch, Nathan and Maltoni, Fabio and Mattelaer, Olivier and Plehn, Tilman and Winterhalder, Ramon},
  title         = {{The MadNIS Reloaded}},
  eprint        = {2311.01548},
  archivePrefix = {arXiv},
  primaryClass  = {hep-ph},
  reportNumber  = {IRMP-CP3-23-56, MCNET-23-12},
  journal       = {arXiv preprint},
  year          = {2023}
}

@inproceedings{Hsieh:2018mld,
  author        = {Hsieh, Ya-Ping and Kavis, Ali and Rolland, Paul and Cevher, Volkan},
  title         = {{Mirrored Langevin Dynamics}},
  booktitle     = {Advances in Neural Information Processing Systems},
  volume        = {31},
  eprint        = {1802.10174},
  archivePrefix = {arXiv},
  primaryClass  = {cs.LG},
  year          = {2018}
}

@incollection{neal2011mcmc,
  title={{MCMC} using {Hamiltonian} dynamics},
  author={Neal, Radford M.},
  booktitle={Handbook of Markov Chain Monte Carlo},
  pages={113--162},
  year={2011},
  publisher={Chapman and Hall/CRC},
  eprint        = {1206.1901},
  archivePrefix = {arXiv},
}

@article{hoffman2014no,
  title={{The No-U-Turn Sampler}: Adaptively Setting Path Lengths in {Hamiltonian Monte Carlo}},
  author={Hoffman, Matthew D. and Gelman, Andrew},
  journal={J. Mach. Learn. Res.},
  volume={15},
  number={1},
  pages={1593--1623},
  year={2014},
  eprint={1111.4246},
  archivePrefix={arXiv},
}

@article{brosse2019tamed,
  title={{The Tamed Unadjusted Langevin Algorithm}},
  author={Brosse, Nicolas and Durmus, Alain and Moulines, {\'E}ric and Sabanis, Sotirios},
  journal={Stochastic Processes and their Applications},
  volume={129},
  number={10},
  pages={3638--3663},
  year={2019},
  publisher={Elsevier},
  eprint={1710.05559},
  archivePrefix={arXiv},
}

@article{roberts1996exponential,
  author  = {Roberts, Gareth O. and Tweedie, Richard L.},
  title   = {{Exponential convergence of Langevin distributions and their discrete approximations}},
  journal = {Bernoulli},
  volume  = {2},
  number  = {4},
  pages   = {341--363},
  year    = {1996}
}

@inproceedings{cheng2018underdamped,
  title={{Underdamped Langevin MCMC}: A Non-Asymptotic Analysis},
  author={Cheng, Xiang and Chatterji, Niladri S and Bartlett, Peter L and Jordan, Michael I},
  booktitle={Conference on Learning Theory},
  pages={300--323},
  year={2018},
  organization={PMLR},
  eprint={1707.03663},
  archivePrefix={arXiv},
}

@article{gelman1997weak,
  title={{Weak Convergence and Optimal Scaling of Random Walk Metropolis Algorithms}},
  author={Gelman, Andrew and Gilks, Walter R and Roberts, Gareth O},
  journal={{The Annals of Applied Probability}},
  volume={7},
  number={1},
  pages={110--120},
  year={1997},
  publisher={Institute of Mathematical Statistics}
}

@article{pasarica2010adaptively,
  title={Adaptively scaling the Metropolis algorithm using expected squared jumped distance},
  author={Pasarica, Cristian and Gelman, Andrew},
  journal={Statistica Sinica},
  pages={343--364},
  year={2010},
  publisher={JSTOR}
}

@article{Roberts:2007adaptive,
  author = {Roberts, Gareth O. and Rosenthal, Jeffrey S.},
  title = {{Coupling and ergodicity of adaptive Markov chain Monte Carlo algorithms}},
  journal = {Journal of Applied Probability},
  volume = {44},
  number = {2},
  pages = {458--475},
  year = {2007},
  doi = {10.1239/jap/1183667414}
}

@article{girolami2011riemann,
  title={{Riemann Manifold Langevin and Hamiltonian Monte Carlo Methods}},
  author={Girolami, Mark and Calderhead, Ben},
  journal={Journal of the Royal Statistical Society Series B: Statistical Methodology},
  volume={73},
  number={2},
  pages={123--214},
  year={2011},
  publisher={Oxford University Press}
}

@article{robbins1951stochastic,
  title={{A Stochastic Approximation Method}},
  author={Robbins, Herbert and Monro, Sutton},
  journal={{The Annals of Mathematical Statistics}},
  pages={400--407},
  year={1951},
  publisher={JSTOR}
}

@article{hu2018stein,
  title={{Stein Neural Sampler}},
  author={Hu, Tianyang and Chen, Zixiang and Sun, Hanxi and Bai, Jincheng and Ye, Mao and Cheng, Guang},
  eprint={1810.03545},
  archivePrefix={arXiv},
  year={2018}
}

@inproceedings{grathwohl2020learning,
  title={{Learning the Stein Discrepancy for Training and Evaluating Energy-Based Models Without Sampling}},
  author={Grathwohl, Will and Wang, Kuan-Chieh and Jacobsen, J{\"o}rn-Henrik and Duvenaud, David and Zemel, Richard},
  booktitle={International Conference on Machine Learning},
  pages={3732--3747},
  year={2020},
  organization={PMLR},
  eprint={2002.05616},
  archivePrefix={arXiv},
}

@inproceedings{gorham2015measuring,
  title={{Measuring Sample Quality with Stein's Method}},
  author={Gorham, Jackson and Mackey, Lester},
  booktitle={Advances in Neural Information Processing Systems},
  volume={28},
  year={2015},
  eprint={1506.03039},
  archivePrefix={arXiv},
}

@article{anastasiou2023stein,
  title={{Stein's Method Meets Computational Statistics: A Review of Some Recent Developments}},
  author={Anastasiou, Andreas and Barp, Alessandro and Briol, Fran{\c{c}}ois-Xavier and Ebner, Bruno and Gaunt, Robert E and Ghaderinezhad, Fatemeh and Gorham, Jackson and Gretton, Arthur and Ley, Christophe and Liu, Qiang and others},
  journal={Statistical Science},
  volume={38},
  number={1},
  pages={120--139},
  year={2023},
  publisher={Institute of Mathematical Statistics},
  eprint={2105.03481},
  archivePrefix={arXiv},
}

@inproceedings{liu2016kernelized,
  title={{A Kernelized Stein Discrepancy for Goodness-of-Fit Tests}},
  author={Liu, Qiang and Lee, Jason and Jordan, Michael},
  booktitle={International Conference on Machine Learning},
  pages={276--284},
  year={2016},
  organization={PMLR},
  eprint={1602.03253},
  archivePrefix={arXiv},
}

@inproceedings{chwialkowski2016kernel,
  title={{A Kernel Test of Goodness of Fit}},
  author={Chwialkowski, Kacper and Strathmann, Heiko and Gretton, Arthur},
  booktitle={International Conference on Machine Learning},
  pages={2606--2615},
  year={2016},
  organization={PMLR},
  eprint={1602.02964},
  archivePrefix={arXiv},
}

@inproceedings{gorham2017measuring,
  title={{Measuring Sample Quality with Kernels}},
  author={Gorham, Jackson and Mackey, Lester},
  booktitle={International Conference on Machine Learning},
  pages={1292--1301},
  year={2017},
  organization={PMLR},
  eprint={1703.01717},
  archivePrefix={arXiv},
}

@article{liu2022estimating,
  title={{Estimating Density Models with Truncation Boundaries Using Score Matching}},
  author={Liu, Song and Kanamori, Takafumi and Williams, Daniel J},
  journal={Journal of Machine Learning Research},
  volume={23},
  number={186},
  pages={1--38},
  year={2022},
  eprint={1910.03834},
  archivePrefix={arXiv},
}

@inproceedings{williams2023approximate,
  title={{Approximate Stein Classes for Truncated Density Estimation}},
  author={Williams, Daniel James and Liu, Song},
  booktitle={International Conference on Machine Learning},
  pages={37066--37090},
  year={2023},
  organization={PMLR},
  eprint={2306.00602},
  archivePrefix={arXiv},
}

@inproceedings{xu2022standardisation,
  title={{Standardisation-Function Kernel Stein Discrepancy: A Unifying View on Kernel Stein Discrepancy Tests for Goodness-of-Fit}},
  author={Xu, Wenkai},
  booktitle={International Conference on Artificial Intelligence and Statistics},
  pages={1575--1597},
  year={2022},
  organization={PMLR},

}

@inproceedings{kingma2014adam,
  author        = {Kingma, Diederik P. and Ba, Jimmy},
  title         = {{Adam: A Method for Stochastic Optimization}},
  booktitle     = {International Conference on Learning Representations},
  eprint        = {1412.6980},
  archivePrefix = {arXiv},
  primaryClass  = {cs.LG},
  year          = {2015}
}

@inproceedings{chewi2021optimal,
  title={{Optimal Dimension Dependence of the Metropolis-Adjusted Langevin Algorithm}},
  author={Chewi, Sinho and Lu, Chen and Ahn, Kwangjun and Cheng, Xiang and Le Gouic, Thibaut and Rigollet, Philippe},
  booktitle={Conference on Learning Theory},
  pages={1260--1300},
  year={2021},
  organization={PMLR},
  eprint={2012.12810},
  archivePrefix={arXiv},
}

@article{Heinrich:2022xfa,
    author = "Heinrich, Lukas and Kagan, Michael",
    title = "{Differentiable Matrix Elements with MadJax}",
    eprint = "2203.00057",
    archivePrefix = "arXiv",
    primaryClass = "hep-ph",
    doi = "10.1088/1742-6596/2438/1/012137",
    journal = "J. Phys. Conf. Ser.",
    volume = "2438",
    number = "1",
    pages = "012137",
    year = "2023"
}

@article{NNPDF:2021njg,
    author = "Ball, Richard D. and others",
    collaboration = "NNPDF",
    title = "{The path to proton structure at 1{\%} accuracy}",
    eprint = "2109.02653",
    archivePrefix = "arXiv",
    primaryClass = "hep-ph",
    reportNumber = "Edinburgh 2021/12, Nikhef-2021-013, TIF-UNIMI-2021-11",
    doi = "10.1140/epjc/s10052-022-10328-7",
    journal = "Eur. Phys. J. C",
    volume = "82",
    number = "5",
    pages = "428",
    year = "2022"
}

@article{Buckley:2014ana,
    author = {Buckley, Andy and Ferrando, James and Lloyd, Stephen and Nordstr{\"o}m, Karl and Page, Ben and R{\"u}fenacht, Martin and Sch{\"o}nherr, Marek and Watt, Graeme},
    title = "{LHAPDF6: parton density access in the LHC precision era}",
    eprint = "1412.7420",
    archivePrefix = "arXiv",
    primaryClass = "hep-ph",
    reportNumber = "GLAS-PPE-2014-05, MCNET-14-29, IPPP-14-111, DCPT-14-222",
    doi = "10.1140/epjc/s10052-015-3318-8",
    journal = "Eur. Phys. J. C",
    volume = "75",
    pages = "132",
    year = "2015"
}

@article{Bothmann:2023siu,
    author = "Bothmann, Enrico and Childers, Taylor and Giele, Walter and Herren, Florian and Hoeche, Stefan and Isaacson, Joshua and Knobbe, Max and Wang, Rui",
    title = "{Efficient phase-space generation for hadron collider event simulation}",
    eprint = "2302.10449",
    archivePrefix = "arXiv",
    primaryClass = "hep-ph",
    reportNumber = "FERMILAB-PUB-23-032-T, MCnet-23-02",
    doi = "10.21468/SciPostPhys.15.4.169",
    journal = "SciPost Phys.",
    volume = "15",
    number = "4",
    pages = "169",
    year = "2023"
}

@misc{bradbury2018jax,
  author       = {Bradbury, James and Frostig, Roy and Hawkins, Peter and Johnson, Matthew James and Katariya, Yash and Leary, Chris and Maclaurin, Dougal and Necula, George and Paszke, Adam and VanderPlas, Jake and Wanderman-Milne, Skye and Zhang, Qiao},
  title        = {{JAX}: composable transformations of {Python}+{NumPy} programs},
  howpublished = {\url{https://github.com/jax-ml/jax}},
  year         = {2018}
}

@article{Bishara:2019iwh,
    author = "Bishara, Fady and Montull, Marc",
    title = "{Machine learning amplitudes for faster event generation}",
    eprint = "1912.11055",
    archivePrefix = "arXiv",
    primaryClass = "hep-ph",
    reportNumber = "DESY 19-232, DESY-19-232",
    doi = "10.1103/PhysRevD.107.L071901",
    journal = "Phys. Rev. D",
    volume = "107",
    number = "7",
    pages = "L071901",
    year = "2023"
}

@article{Badger:2020uow,
    author = "Badger, Simon and Bullock, Joseph",
    title = "{Using neural networks for efficient evaluation of high multiplicity scattering amplitudes}",
    eprint = "2002.07516",
    archivePrefix = "arXiv",
    primaryClass = "hep-ph",
    reportNumber = "IPPP/20/5",
    doi = "10.1007/JHEP06(2020)114",
    journal = "JHEP",
    volume = "06",
    pages = "114",
    year = "2020"
}

@article{Aylett-Bullock:2021hmo,
    author = "Aylett-Bullock, Joseph and Badger, Simon and Moodie, Ryan",
    title = "{Optimising simulations for diphoton production at hadron colliders using amplitude neural networks}",
    eprint = "2106.09474",
    archivePrefix = "arXiv",
    primaryClass = "hep-ph",
    doi = "10.1007/JHEP08(2021)066",
    journal = "JHEP",
    volume = "08",
    pages = "066",
    year = "2021"
}

@article{Maitre:2021uaa,
    author = "Ma{\^\i}tre, Daniel and Truong, Henry",
    title = "{A factorisation-aware Matrix element emulator}",
    eprint = "2107.06625",
    archivePrefix = "arXiv",
    primaryClass = "hep-ph",
    reportNumber = "IPPP/21/11",
    doi = "10.1007/JHEP11(2021)066",
    journal = "JHEP",
    volume = "11",
    pages = "066",
    year = "2021"
}

@article{Danziger:2021eeg,
    author = "Danziger, Katharina and Jan{\ss}en, Timo and Schumann, Steffen and Siegert, Frank",
    title = "{Accelerating Monte Carlo event generation -- rejection sampling using neural network event-weight estimates}",
    eprint = "2109.11964",
    archivePrefix = "arXiv",
    primaryClass = "hep-ph",
    reportNumber = "MCNET-21-13",
    doi = "10.21468/SciPostPhys.12.5.164",
    journal = "SciPost Phys.",
    volume = "12",
    pages = "164",
    year = "2022"
}

@article{Badger:2022hwf,
    author = "Badger, Simon and Butter, Anja and Luchmann, Michel and Pitz, Sebastian and Plehn, Tilman",
    title = "{Loop amplitudes from precision networks}",
    eprint = "2206.14831",
    archivePrefix = "arXiv",
    primaryClass = "hep-ph",
    doi = "10.21468/SciPostPhysCore.6.2.034",
    journal = "SciPost Phys. Core",
    volume = "6",
    pages = "034",
    year = "2023"
}

@article{Janssen:2023ahv,
    author = "Jan{\ss}en, Timo and Ma{\^\i}tre, Daniel and Schumann, Steffen and Siegert, Frank and Truong, Henry",
    title = "{Unweighting multijet event generation using factorisation-aware neural networks}",
    eprint = "2301.13562",
    archivePrefix = "arXiv",
    primaryClass = "hep-ph",
    reportNumber = "MCNET-23-01, IPPP/23/04",
    doi = "10.21468/SciPostPhys.15.3.107",
    journal = "SciPost Phys.",
    volume = "15",
    number = "3",
    pages = "107",
    year = "2023"
}

@article{Maitre:2023dqz,
    author = "Ma{\^\i}tre, D. and Truong, H.",
    title = "{One-loop matrix element emulation with factorisation awareness}",
    eprint = "2302.04005",
    archivePrefix = "arXiv",
    primaryClass = "hep-ph",
    reportNumber = "IPPP/23/06",
    doi = "10.1007/JHEP05(2023)159",
    journal = "JHEP",
    volume = "05",
    pages = "159",
    year = "2023"
}

@article{Bahl:2024gyt,
    author = "Bahl, Henning and Elmer, Nina and Favaro, Luigi and Haussmann, Manuel and Plehn, Tilman and Winterhalder, Ramon",
    title = "{Accurate surrogate amplitudes with calibrated uncertainties}",
    eprint = "2412.12069",
    archivePrefix = "arXiv",
    primaryClass = "hep-ph",
    doi = "10.21468/SciPostPhysCore.8.4.073",
    journal = "SciPost Phys. Core",
    volume = "8",
    pages = "073",
    year = "2025"
}

@article{Brehmer:2024yqw,
    author = "Brehmer, Johann and Bres{\'o}, V{\'\i}ctor and de Haan, Pim and Plehn, Tilman and Qu, Huilin and Spinner, Jonas and Thaler, Jesse",
    title = "{A Lorentz-equivariant transformer for all of the LHC}",
    eprint = "2411.00446",
    archivePrefix = "arXiv",
    primaryClass = "hep-ph",
    reportNumber = "MIT-CTP/5802",
    doi = "10.21468/SciPostPhys.19.4.108",
    journal = "SciPost Phys.",
    volume = "19",
    number = "4",
    pages = "108",
    year = "2025"
}

@article{Breso-Pla:2024pda,
    author = "Bres{\'o}-Pla, V{\'\i}ctor and Heinrich, Gudrun and Magerya, Vitaly and Olsson, Anton",
    title = "{Interpolating amplitudes}",
    eprint = "2412.09534",
    archivePrefix = "arXiv",
    primaryClass = "hep-ph",
    reportNumber = "CERN-TH-2024-211, KA-TP-23-2024, P3H-24-092",
    doi = "10.21468/SciPostPhys.19.5.123",
    journal = "SciPost Phys.",
    volume = "19",
    number = "5",
    pages = "123",
    year = "2025"
}

@article{Herrmann:2025nnz,
    author = "Herrmann, Tim and Jan{\ss}en, Timo and Schenker, Mathis and Schumann, Steffen and Siegert, Frank",
    title = "{Accelerating multijet-merged event generation with neural network matrix element surrogates}",
    eprint = "2506.06203",
    archivePrefix = "arXiv",
    primaryClass = "hep-ph",
    reportNumber = "MCNET-25-12",
    doi = "10.21468/SciPostPhys.20.3.071",
    journal = "SciPost Phys.",
    volume = "20",
    pages = "071",
    year = "2026"
}

@article{Bahl:2025xvx,
    author = "Bahl, Henning and Elmer, Nina and Plehn, Tilman and Winterhalder, Ramon",
    title = "{Amplitude Uncertainties Everywhere All at Once}",
    eprint = "2509.00155",
    archivePrefix = "arXiv",
    primaryClass = "hep-ph",
    reportNumber = "TIF-UNIMI-2025-17",
    doi = "10.21468/SciPostPhys.20.3.083",
    journal = "SciPost Phys.",
    volume = "20",
    pages = "083",
    year = "2026"
}

@article{Favaro:2025pgz,
    author = "Favaro, Luigi and Gerhartz, Gerrit and Hamprecht, Fred A. and Lippmann, Peter and Pitz, Sebastian and Plehn, Tilman and Qu, Huilin and Spinner, Jonas",
    title = "{Lorentz-Equivariance without Limitations}",
    eprint = "2508.14898",
    archivePrefix = "arXiv",
    primaryClass = "hep-ph",
    month = "8",
    year = "2025"
}

@article{Villadamigo:2025our,
    author = "Villadamigo, Javier Mari{\~n}o and Frederix, Rikkert and Plehn, Tilman and Vitos, Timea and Winterhalder, Ramon",
    title = "{FASTColor -- Full-color Amplitude Surrogate Toolkit for QCD}",
    eprint = "2509.07068",
    archivePrefix = "arXiv",
    primaryClass = "hep-ph",
    reportNumber = "TIF-UNIMI-2025-18",
    month = "9",
    year = "2025"
}

@article{Beccatini:2025tpk,
    author = "Beccatini, Luca and Maltoni, Fabio and Mattelaer, Olivier and Winterhalder, Ramon",
    title = "{Amplitude Surrogates for Multi-Jet Processes}",
    eprint = "2512.11036",
    archivePrefix = "arXiv",
    primaryClass = "hep-ph",
    reportNumber = "TIF-UNIMI-2025-26, IRMP-CP3-25-44",
    month = "12",
    year = "2025"
}

@article{Bahl:2026jvt,
    author = "Bahl, Henning and Bres{\'o}-Pla, Victor and Butter, Anja and Ramirez, Joaqu{\'\i}n Iturriza",
    title = "{Scaling laws for amplitude surrogates}",
    eprint = "2601.13308",
    archivePrefix = "arXiv",
    primaryClass = "hep-ph",
    month = "1",
    year = "2026"
}

@article{Bahl:2026qaf,
    author = "Bahl, Henning and Braun, Jens and Heinrich, Gudrun and Plehn, Tilman and Revelli, Rebecca",
    title = "{How to Trust Learned Loop Amplitudes}",
    eprint = "2601.00950",
    archivePrefix = "arXiv",
    primaryClass = "hep-ph",
    month = "1",
    year = "2026"
}

@article{Carrazza:2021zug,
    author = "Carrazza, Stefano and Cruz-Martinez, Juan and Rossi, Marco and Zaro, Marco",
    title = "{MadFlow: towards the automation of Monte Carlo simulation on GPU for particle physics processes}",
    eprint = "2105.10529",
    archivePrefix = "arXiv",
    primaryClass = "physics.comp-ph",
    reportNumber = "TIF-UNIMI-2021-3",
    doi = "10.1051/epjconf/202125103022",
    journal = "EPJ Web Conf.",
    volume = "251",
    pages = "03022",
    year = "2021"
}

@article{Carrazza:2021gpx,
    author = "Carrazza, Stefano and Cruz-Martinez, Juan and Rossi, Marco and Zaro, Marco",
    title = "{MadFlow: automating Monte Carlo simulation on GPU for particle physics processes}",
    eprint = "2106.10279",
    archivePrefix = "arXiv",
    primaryClass = "physics.comp-ph",
    reportNumber = "TIF-UNIMI-2021-9",
    doi = "10.1140/epjc/s10052-021-09443-8",
    journal = "Eur. Phys. J. C",
    volume = "81",
    number = "7",
    pages = "656",
    year = "2021"
}

@article{Heimel:2024wph,
    author = "Heimel, Theo and Mattelaer, Olivier and Plehn, Tilman and Winterhalder, Ramon",
    title = "{Differentiable MadNIS-Lite}",
    eprint = "2408.01486",
    archivePrefix = "arXiv",
    primaryClass = "hep-ph",
    reportNumber = "IRMP-CP3-24-23",
    doi = "10.21468/SciPostPhys.18.1.017",
    journal = "SciPost Phys.",
    volume = "18",
    number = "1",
    pages = "017",
    year = "2025"
}

@article{Heimel:2026hgp,
    author = "Heimel, Theo and Mattelaer, Olivier and Winterhalder, Ramon",
    title = "{MadSpace -- Event Generation for the Era of GPUs and ML}",
    eprint = "2602.06895",
    archivePrefix = "arXiv",
    primaryClass = "hep-ph",
    reportNumber = "MCNET-26-01, IRMP-CP3-26-04, TIF-UNIMI-2026-1",
    month = "2",
    year = "2026"
}

@article{Buckley:2011ms,
    author = "Buckley, Andy and others",
    title = "{General-purpose event generators for LHC physics}",
    eprint = "1101.2599",
    archivePrefix = "arXiv",
    primaryClass = "hep-ph",
    reportNumber = "CAVENDISH-HEP-10-21, CERN-PH-TH-2010-298, DCPT-10-202, IPPP-10-101, KA-TP-40-2010, LU-TP-10-28, MAN-HEP-2010-23, SLAC-PUB-14333, HD-THEP-10-24, MCNET-11-01",
    doi = "10.1016/j.physrep.2011.03.005",
    journal = "Phys. Rept.",
    volume = "504",
    pages = "145--233",
    year = "2011"
}

@article{Narain:2022qud,
    author = "Narain, Meenakshi and others",
    title = "{The Future of US Particle Physics - The Snowmass 2021 Energy Frontier Report}",
    eprint = "2211.11084",
    archivePrefix = "arXiv",
    primaryClass = "hep-ex",
    reportNumber = "FERMILAB-FN-1219-PPD-T",
    doi = "10.2172/1908199",
    month = "11",
    year = "2022"
}

@article{HEPSoftwareFoundation:2017ggl,
    author = "Albrecht, Johannes and others",
    collaboration = "HEP Software Foundation",
    title = "{A Roadmap for HEP Software and Computing R{\&}D for the 2020s}",
    eprint = "1712.06982",
    archivePrefix = "arXiv",
    primaryClass = "physics.comp-ph",
    reportNumber = "HSF-CWP-2017-01, HSF-CWP-2017-001, FERMILAB-PUB-17-607-CD",
    doi = "10.1007/s41781-018-0018-8",
    journal = "Comput. Softw. Big Sci.",
    volume = "3",
    number = "1",
    pages = "7",
    year = "2019"
}

@article{HSFPhysicsEventGeneratorWG:2020gxw,
    author = "Amoroso, Simone and others",
    editor = "Valassi, Andrea and Yazgan, Efe and McFayden, Josh",
    collaboration = "HSF Physics Event Generator WG",
    title = "{Challenges in Monte Carlo Event Generator Software for High-Luminosity LHC}",
    eprint = "2004.13687",
    archivePrefix = "arXiv",
    primaryClass = "hep-ph",
    reportNumber = "CERN-LPCC-2020-002, FERMILAB-PUB-20-183-SCD-T, MCNET-20-15",
    doi = "10.1007/s41781-021-00055-1",
    journal = "Comput. Softw. Big Sci.",
    volume = "5",
    number = "1",
    pages = "12",
    year = "2021"
}

@book{EuropeanStrategyGroup:2020pow,
    author = "{European Strategy Group}",
    title = "{2020 Update of the European Strategy for Particle Physics}",
    reportNumber = "CERN-ESU-013, CERN-ESU-015",
    doi = "10.17181/ESU2020",
    isbn = "978-92-9083-575-2",
    publisher = "CERN Council",
    address = "Geneva",
    year = "2020"
}

@book{ZurbanoFernandez:2020cco,
    author = "Zurbano Fernandez, I. and others",
    title = "{High-Luminosity Large Hadron Collider (HL-LHC): Technical design report}",
    series = "CERN Yellow Reports: Monographs",
    number = "10/2020",
    publisher = "CERN",
    address = "Geneva",
    doi = "10.23731/CYRM-2020-0010",
    month = "12",
    year = "2020"
}

@article{FCC:2018evy,
    author = "Abada, A. and others",
    collaboration = "FCC",
    title = "{FCC-ee: The Lepton Collider}: {Future Circular Collider Conceptual Design Report Volume 2}",
    reportNumber = "CERN-ACC-2018-0057",
    doi = "10.1140/epjst/e2019-900045-4",
    journal = "Eur. Phys. J. ST",
    volume = "228",
    number = "2",
    pages = "261--623",
    year = "2019"
}

@article{FCC:2018vvp,
    author = "Abada, A. and others",
    collaboration = "FCC",
    title = "{FCC-hh: The Hadron Collider}: {Future Circular Collider Conceptual Design Report Volume 3}",
    reportNumber = "CERN-ACC-2018-0058",
    doi = "10.1140/epjst/e2019-900087-0",
    journal = "Eur. Phys. J. ST",
    volume = "228",
    number = "4",
    pages = "755--1107",
    year = "2019"
}

@article{vanBeekveld:2026uxl,
    author = {van Beekveld, Melissa and Bothmann, Enrico and Buckley, Andy and G{\"u}tschow, Christian and Skands, Peter and Winterhalder, Ramon},
    title = "{The Monte Carlo Ecosystem in High-Energy Physics: A Primer}",
    eprint = "2605.16036",
    archivePrefix = "arXiv",
    primaryClass = "hep-ph",
    reportNumber = "MCNET-26-11, TIF-UNIMI-2026-7",
    month = "5",
    year = "2026"
}

@article{Lepage:1977sw,
    author = "Lepage, G. Peter",
    title = "{A New Algorithm for Adaptive Multidimensional Integration}",
    reportNumber = "SLAC-PUB-1839-REV, SLAC-PUB-1839",
    doi = "10.1016/0021-9991(78)90004-9",
    journal = "J. Comput. Phys.",
    volume = "27",
    pages = "192",
    year = "1978"
}

@article{Lepage:2020tgj,
    author = "Lepage, G. Peter",
    title = "{Adaptive multidimensional integration: VEGAS enhanced}",
    eprint = "2009.05112",
    archivePrefix = "arXiv",
    primaryClass = "physics.comp-ph",
    doi = "10.1016/j.jcp.2021.110386",
    journal = "J. Comput. Phys.",
    volume = "439",
    pages = "110386",
    year = "2021"
}

@article{Kleiss:1994qy,
    author = "Kleiss, Ronald and Pittau, Roberto",
    title = "{Weight optimization in multichannel Monte Carlo}",
    eprint = "hep-ph/9405257",
    archivePrefix = "arXiv",
    reportNumber = "NIKHEF-H-94-17, INLO-PUB-4-94",
    doi = "10.1016/0010-4655(94)90043-4",
    journal = "Comput. Phys. Commun.",
    volume = "83",
    pages = "141--146",
    year = "1994"
}

@article{Ohl:1998jn,
    author = "Ohl, Thorsten",
    title = "{Vegas revisited: Adaptive Monte Carlo integration beyond factorization}",
    eprint = "hep-ph/9806432",
    archivePrefix = "arXiv",
    reportNumber = "IKDA-98-15",
    doi = "10.1016/S0010-4655(99)00209-X",
    journal = "Comput. Phys. Commun.",
    volume = "120",
    pages = "13--19",
    year = "1999"
}

@article{Maltoni:2002qb,
    author = "Maltoni, Fabio and Stelzer, Tim",
    title = "{MadEvent: Automatic event generation with MadGraph}",
    eprint = "hep-ph/0208156",
    archivePrefix = "arXiv",
    doi = "10.1088/1126-6708/2003/02/027",
    journal = "JHEP",
    volume = "02",
    pages = "027",
    year = "2003"
}

@article{Krauss:2001iv,
    author = "Krauss, F. and Kuhn, R. and Soff, G.",
    title = "{AMEGIC++ 1.0: A Matrix element generator in C++}",
    eprint = "hep-ph/0109036",
    archivePrefix = "arXiv",
    reportNumber = "CAVENDISH-HEP-01-11",
    doi = "10.1088/1126-6708/2002/02/044",
    journal = "JHEP",
    volume = "02",
    pages = "044",
    year = "2002"
}

@article{Papadopoulos:2000tt,
    author = "Papadopoulos, Costas G.",
    title = "{PHEGAS: A Phase space generator for automatic cross-section computation}",
    eprint = "hep-ph/0007335",
    archivePrefix = "arXiv",
    doi = "10.1016/S0010-4655(01)00163-1",
    journal = "Comput. Phys. Commun.",
    volume = "137",
    pages = "247--254",
    year = "2001"
}

@article{Kilian:2007gr,
    author = "Kilian, Wolfgang and Ohl, Thorsten and Reuter, Jurgen",
    title = "{WHIZARD: Simulating Multi-Particle Processes at LHC and ILC}",
    eprint = "0708.4233",
    archivePrefix = "arXiv",
    primaryClass = "hep-ph",
    reportNumber = "DESY-11-126, EDINBURGH-2010-36, FR-PHENO-2010-037, SI-HEP-2010-18",
    doi = "10.1140/epjc/s10052-011-1742-y",
    journal = "Eur. Phys. J. C",
    volume = "71",
    pages = "1742",
    year = "2011"
}

@article{Jadach:2002kn,
    author = "Jadach, S.",
    title = "{Foam: A General purpose cellular Monte Carlo event generator}",
    eprint = "physics/0203033",
    archivePrefix = "arXiv",
    reportNumber = "CERN-TH-2002-059",
    doi = "10.1016/S0010-4655(02)00755-5",
    journal = "Comput. Phys. Commun.",
    volume = "152",
    pages = "55--100",
    year = "2003"
}

@article{Gleisberg:2008fv,
    author = "Gleisberg, Tanju and Hoeche, Stefan",
    title = "{Comix, a new matrix element generator}",
    eprint = "0808.3674",
    archivePrefix = "arXiv",
    primaryClass = "hep-ph",
    reportNumber = "SLAC-PUB-13232, IPPP-08-31, DCPT-08-62, MCNET-08-08",
    doi = "10.1088/1126-6708/2008/12/039",
    journal = "JHEP",
    volume = "12",
    pages = "039",
    year = "2008"
}

@article{Hoche:2019flt,
    author = {H{\"o}che, Stefan and Prestel, Stefan and Schulz, Holger},
    title = "{Simulation of Vector Boson Plus Many Jet Final States at the High Luminosity LHC}",
    eprint = "1905.05120",
    archivePrefix = "arXiv",
    primaryClass = "hep-ph",
    reportNumber = "FERMILAB-PUB-19-192-T, LU-TP 19-14, MCNET-19-09",
    doi = "10.1103/PhysRevD.100.014024",
    journal = "Phys. Rev. D",
    volume = "100",
    number = "1",
    pages = "014024",
    year = "2019"
}

\end{document}